\documentclass[12pt]{article}
\usepackage{amsmath,amsfonts, amssymb,graphicx,geometry,authblk,color,soul,comment,hyperref,subcaption} 
\usepackage{bm, algorithm, algpseudocode}
\DeclareMathOperator{\argmin}{argmin}

\newcommand{\la}{\left\langle}
\newcommand{\ra}{\right\rangle}

\title{Learning a potential formulation for rate-and-state friction}
\author{Shengduo Liu, Kaushik Bhattacharya, Nadia Lapusta}
\affil{California Institute of Technology}
\begin{document}
\maketitle

\begin{center}
{\it Dedicated to Michael Ortiz, an inspiring mentor, colleague and friend.}
\end{center}

\begin{abstract}
Empirical rate-and-state friction laws are widely used in geophysics and engineering to simulate interface slip. They postulate that the friction coefficient depends on the local slip rate and a state variable that reflects the history of slip.  Depending on the parameters, rate-and-state friction can be either rate-strengthening, leading to steady slip, or rate-weakening, leading to unsteady stick-slip behavior modeling earthquakes.   Rate-and-state friction does not have a potential or variational formulation, making implicit solution approaches difficult and implementation numerically expensive.  In this work, we propose a potential formulation for the rate-and-state friction. We formulate the potentials as neural networks and train them so that the resulting behavior emulates the empirical rate-and-state friction.  We show that this potential formulation enables implicit time discretization leading to efficient numerical implementation.
\end{abstract}

%%%%%%%%%%%%%%%%%%%%%%%%%%%%%%%%%%%%%%%%%
\section{Introduction}
\label{sec:introduction}

It has long been recognized from experiments and theoretical ideas that the friction strength of interfaces and thin shear layers (or faults) in rocks and other materials depends both on the history of slip and on slip velocity (or rate) (e.g., \cite{Rabinowicz1951, Rabinowicz1958Intrinsic,Ida1972,Scholz1972,palmer_growth_1973,Dieterich1978}). Rate-and-state friction laws were formulated based on laboratory experiments \cite{dieterich_modeling_1979,dieterich_potential_1981,ruina_slip_1983}), and combine the dependence of frictional strength on rate (or slip velocity) and, through a state variable, history of slip. The empirical rate-and-state formulations have been used to reproduce, both qualitatively and quantitatively, a number of earthquake-source observations, including earthquake nucleation, interseismic creep, postseismic slip, aftershock sequences, scaling of small repeating earthquakes, and slow slip events \cite{marone_laboratory-derived_1998, Dieterich2007,  ChenLapusta2009, Kaneko2010Towards}.

The numerical simulations of problems involving rate-and-state friction are expensive, especially in the rate-weakening setting that can lead to large accelerations and stick-slip events.  This is made more difficult by the fact that the empirical rate-and-state friction laws do not have a potential formulation, as we discuss later in this work.  Therefore it is not possible to formulate an implicit time discretization as a variational problem and implicit-time updates do not necessarily converge.  To enable further efficiency and stability of numerical simulations, it is desirable to have a friction formulation that matches the empirical rate-and-state friction, and hence key features of the experimental data, but enables stable implicit time discretization.  This is the overarching goal of this study.

There is a long history of the use of internal or state variable theories in continuum mechanics (e.g., \cite{rice1971inelastic,gurtin_cont_mech}) (and more broadly the use of order parameters in physics) to develop the architecture for our machine-learnt approximation.  These theories postulate that all prior history can be represented through state or internal variables that evolve with the deformation.  The internal and state variable theories inspired the introduction of recurrent neural operators (RNOs) as a neural network representation of history dependent phenomena \cite{BurigedeEric2023, BurigedeMarkovian2023}.  Here, the material and evolution laws are represented as neural networks whose parameters as well as the underlying state variables are learnt from the data.  These are continuous in time formulations, and can be used at any time-discretization.  In many theories, in fact most cases, the evolution law may be represented in terms of a dissipation potential.  Consequently, the implicit-time discretization of the problem leads to a variational problem that enables efficient and stable solution (e.g,, \cite{ortiz_1999,miehe_2002,mielke_2005}).  Recently, RNOs have been extended to the potential setting \cite{Raj}.

In this work, we propose a potential formulation of friction laws that leads to a convex variational problem for the implicit time-discretization, but emulates the behavior of the empirical rate-and-state friction.  Our approach is inspired by RNOs, in that we postulate that the potentials in the formulation are deep neural networks that are trained with data provided by the empirical rate-and-state friction law.  This requires care since rate-and-state friction laws is an unusual state variable theory.  

Recall that in the simplest model of Coulomb friction, the coefficient of static friction coefficient is typically higher than that of dynamic friction coefficient.  This leads to unstable sliding.  In a more careful rate-and-state formulation, the friction coefficient is postulated to depend on the slip rate.  However, in the rate-weakening setting, the friction coefficient can drop with increasing slip rate.  This too is unstable and can lead to ill-posed problems \cite{rice_stability_1983}.  Indeed, rate-weakening would lead to a non-convex dissipation potential that depends on slip-rate.  The state variable stabilizes the situation but still enables dynamic bursts by increasing the friction coefficient in a transient manner when the sliding rate increases suddenly.  In other words, the state variable acts in a manner that is related to the slip acceleration.  This means that we have to introduce two dissipation potentials, one that depends on slip rate and one that depends on the internal variable evolution, but which act at different powers of time-step in a discrete formulation.

After briefly recalling rate-and-state friction in Section \ref{sec:rs}, we introduce our potential formulation in Section \ref{sec:formulation}.  We introduce the neural approximation and train it to the data obtained using empirical rate-and-state laws in Section \ref{sec:NN}.  Specifically, we show that a potential formulation with a single internal variable is able to reproduce the behavior of the empirical rate-and-state laws within an error that is smaller than the error of matching the empirical rate-and-state law to experimental data.  We then apply the potential formulation to a spring-slider system in Section \ref{sec:sss}.  We show that an implicit-time update of this system results in a minimization problem and derive conditions under which it is a convex minimization problem.  We also demonstrate that the potential formulation leads to significantly improved numerical performance.  We conclude in Section \ref{sec:conclusions}.

\section{Rate-and-State Friction}
\label{sec:rs}

The Coulomb formulation of friction states that the shear traction $\tau$ a material point $X$ on a frictional interface is related to the applied normal traction $\sigma$ through friction coefficient $f$.  It is widely recognized that the friction coefficient depends on the slip history of that location $\left\{x(X, t') : t' \in [0, t]\right\}$, i.e., 
\begin{align}
    \tau(t, X) = \sigma(t, X) f\left(\left\{x(X, t') : t' \in [0, t]\right\}\right) \label{eq:generalFric}. 
\end{align}
\noindent We denote slip by $x$, as it is a common notation in the spring-slider model considered later, where the shear displacement $x$ from the zero position coincides with slip.

Rate-and-state formulations of friction postulate that the dependency on slip history can be described through the current slip rate, $V = \dot{x}(X, t)$ and a state variable $\theta(X, t)$. 
A widely used empirical rate-and-state formulation, inspired by experimental observations \cite{dieterich_modeling_1979, marone_laboratory-derived_1998, ruina_slip_1983}, is:
\begin{align}
    f^{RS}(X, t) = f_* + a \log\left(\frac{V(X, t)}{V_*}\right) + b \log\left(\frac{V_* \theta(X, t)} {D_{RS}}\right) \label{eq:fRS}, 
\end{align}
where $f_*$ is reference friction coefficient, 
$V_*$ is reference slip rate, 
$D_{RS}$ is characteristic slip distance, 
$a, b$ are dimensionless rate-and-state parameters, 
and $\theta$ is a state variable that evolves with time. 
The evolution of $\theta$ is given by \cite{dieterich_modeling_1979, ruina_slip_1983}:
\begin{align}
    \dot{\theta}(X, t) = 1 - \frac{V(X, t) \theta(X, t)}{D_{RS}} \label{eq:AgeingLaw}. 
\end{align}
Other evolution laws for the state-variable evolution have been proposed \cite{dieterich_modeling_1979, ruina_slip_1983, PerrinRiceZheng1995} and the approach for developing the potential formulation presented in this work should apply to them as well. Note that this formulation is local and does not explicitly depend on $X$; so we suppress $X$ in the notation without ambiguity. 

At steady state ($\dot{\theta} = \dot{V} = 0$), this law reduces to
\begin{align}
    f_{ss}^{RS} = f_* + (a - b) \log \left(\frac{V}{V_*}\right) \label{eq:fRSss}. 
\end{align}
If $a - b > 0$, 
the steady rate-and-state friction coefficient $f^{RS}$ increases as slip rate $V$ increases, 
and the friction is rate-strengthening. 
If $a - b < 0$, 
the friction is rate-weakening. 
Rate-weakening rate-and-state friction can lead to large accelerations and can simulate dynamic slip of earthquake rupture \cite{dieterich_modeling_1979, 
 marone_laboratory-derived_1998, ruina_slip_1983,rice_stability_1983, scholz_2019}. 

It can be easily verified that there is no potential associated with this empirical rate-and-state formulation, i.e., one cannot find scalar potential functions whose gradients would yield both the friction coefficient $f^{RS}$ and the evolution law of $\theta$. Briefly, this is due to the cross term $V\theta$ in the evolution law.

%%%%%%%%%%%%%%%%%%%%%%%%%%%%%%%%%%%%%%%%%
\section{Potential Formulation}
\label{sec:formulation}

We propose a potential formulation of the general history-dependent friction given by (\ref{eq:generalFric}).  We postulate that the history of slip  can be modeled as a dependence on a $D$-dimensional vector of internal variables\footnote{We could also call these state variables, but reserve the term for the state variable of rate-and-state friction to avoid confusion.} $\bm{\xi} \in \mathbb{R}^D$.   
We introduce two potentials,  $\widetilde D^\dagger(\dot{x}, \bm{\xi})$ and $D(\dot{\bm{\xi}})$, whose derivatives give rise to the friction coefficient and evolution law of $\bm{\xi}$, i.e., 
\begin{align}
    &f^{P}(\dot{x}, \boldsymbol{\xi}) =  \frac{\partial \widetilde D^\dagger}{\partial \dot{x}}(\dot{x}, \bm{\xi}) \label{eq:fpot}, \\
    &\frac{d D}{d \dot{\boldsymbol{\xi}}}(\dot{\bm{\xi}}) + \frac{\partial D^\dagger}{\partial \boldsymbol{\xi}}(\dot{x}, \bm{\xi}) = 0 \label{eq:evolutionXi},  
\end{align}
where $f^{P}$ is the friction coefficient of the potential rate-and-state formulation,  $x$ is local slip, $V = \dot{x}$ is local slip rate. 

In practice, it is useful to make two modifications.  First, we write
\begin{equation}
\widetilde D^\dagger = f_* V + D^\dagger
\end{equation}
to account for the reference friction $f_*$.  Second, it is useful to work with the dual formulation of expression (\ref{eq:evolutionXi}).
Let $D^*(\bm{d})$ be the Legendre transform of $D(\bm{\dot{\xi}})$, 
\begin{align}
    D^*\left(\bm{d}\right) &= \sup_{\bm{\dot{\xi}} \in \mathbb{R}^d} \left\{\la \bm{d}, \bm{\dot{\xi}} \ra -D(\dot{\bm{\xi}})\right\}. \label{eq:LegendreDstar}
\end{align} 
The advantage of using $D^*$ instead of $D$ is that, instead of solving the non-linear equation for $\dot{\bm{\xi}}$ given by (\ref{eq:evolutionXi}), 
one can compute $\dot{\bm{\xi}}$ by:
\begin{align}
    \dot{\bm{\xi}} &= \frac{d \Tilde{D}^*}{d \bm{d}}\left(-\frac{\partial \Tilde{D}^\dagger}{\partial \bm{\xi}}\right) \label{eq:ComputeXiDot}, 
\end{align}
if $D\left(\bm{\dot{\xi}}\right)$ is convex in $\bm{\dot{\xi}}$.

%%%%%%%%%%%%%%%%%%%%%%%%%%%%%%%%%%%%%%%%%
\section{Learning potentials that mimic empirical rate-and-state friction}
\label{sec:NN}

%%%%%%%%%%%%%%%%%%%%%%%%%%%%%%%%%%%%%%%%%
\subsection{Neural approximation of the potentials}

\begin{figure}
    \centering
    \includegraphics[width=0.8\textwidth]{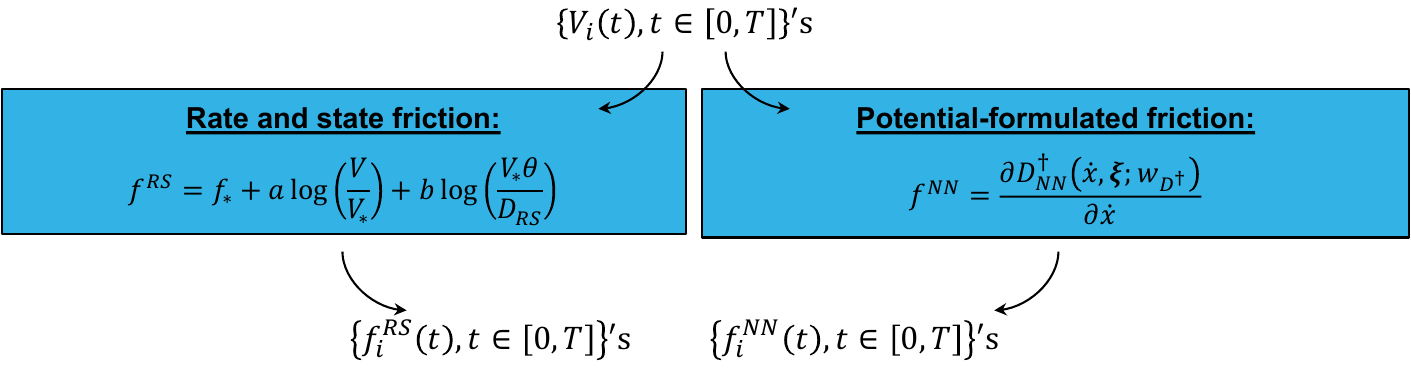}
    \caption{Training of $D^\dagger$ and $D$ through fitting $f^{NN}$ to $f^{RS}$.}
    \label{fig:training}
\end{figure}

We seek to learn the potentials $D^\dagger, D^*$ as well as the internal variable $\bm{\xi}$ such that the proposed variational formulation (\ref{eq:fpot}, \ref{eq:ComputeXiDot}) emulates the behavior of the empirical rate-and-state formulation (\ref{eq:fRS}-\ref{eq:AgeingLaw}).  Specifically, we assume that the potentials $D^\dagger, D^*$ are described by hyper-parametrized deep neural networks,
\begin{align}
     D^\dagger(\dot{x}, \bm{\xi}) \approx D^\dagger_{NN}(\dot{x}, \bm{\xi}; w_{D^\dagger}), \quad
    D^*\left(\dot{\bm{d}}\right) \approx D^*_{NN}\left(\dot{\bm{d}}, w_{D^*}\right)\label{eq:NNpotentials},  
\end{align}
with parameters $w = \{w_{D^\dagger},w_{D^*}\}$
and we train the networks using data generated from the empirical rate-and-state formulation.  We select a set of velocity histories, $\{V_i (t), t\in[0,T]\}_{i=1}^N$, and calculate the resulting friction coefficients, $\{f_i^{RS}\}_{i=1}^N$, from the empirical rate-and-state laws (\ref{eq:fRS},\ref{eq:AgeingLaw}) and $\{f_i^{NN}]\}_{i=1}^N$ from the proposed model (\ref{eq:fpot},\ref{eq:ComputeXiDot},\ref{eq:NNpotentials}).  We then find the internal variables and the parameters $w$ that minimize the relative $L_p$ norm:
\begin{align}
L =  \frac{1}{N} \sum_{i=1}^N \frac{\|f^{NN}_i(t) - f^{RS}_i(t)\|_{L_p}}{\|f^{RS}_i(t)\|_{L_p}} \label{eq:relativeLpError}
\end{align}
of the difference between the two friction coefficients.  This is shown schematically in Figure~\ref{fig:training}.  Note that the architecture (\ref{eq:fpot}, \ref{eq:ComputeXiDot},\ref{eq:NNpotentials}) is a recurrent neural operator (RNO) following \cite{BurigedeEric2023, BurigedeMarkovian2023}.  An important aspect of this approach is that the state variables are not defined a priori, but discovered from the data during training.

We implement this in the environment PyTorch \cite{paszke2019pytorch}.  We use the Optuna \cite{akiba2019optuna} optimization package for hyper-parameter tuning including 
the depth of the Neural Network, 
the number of neurons within each layer, 
the value of $p$ for the $L_p$ norm, 
as well as the batch size in the training dataset. 
The detailed algorithm of the training process can be found in Algorithm~\ref{alg:TrainingOneEpoch}. 

\begin{algorithm}[t]
\caption{Training $D_{NN}^\dagger(\dot{x}, \bm{\xi}; w_{D^\dagger})$ and $D_{NN}^*(\dot{\bm{d}}; w_D)$  }\label{alg:TrainingOneEpoch}
\begin{algorithmic}
%% Setting parameters
% Input space
\Require Training data $\left\{V_{i}(t), f^{RS}_i(t) : t \in [0, T]\right\}_{i = 0}^N = \cup_{b=1}^B \left\{V_{i}(t), f^{RS}_i(t) : t \in [0, T]\right\}_{i = 0}^{N_b}$ divided into $B$ batches
\Require $N_{epochs}$
%% Algorithm begins
\State $epoch = 1$
\While{$epoch\le N_{epochs}$}
\State{$b=1$}
\While{$b\le B$}
    \For {$i \in \{0, 1, ..., N_b\}$} %\Comment{In practical sequences are passed in batches.}
        \State Fix $w_{D^\dagger}$, $w_D$
        \For {$n = 1, 2, ..., N^{(i)}$} %\Comment{$N^{(i)}$ is number of time steps of sequence $i$. }
            \State $\xi_n \gets \xi_{n-1} + (t_n-t_{n-1}) \dot{\bm{\xi}}_{n-1}$
            \State $f^{NN}_n \gets \frac{\partial W_{NN}}{\partial x}(x_n) + \frac{\partial D^\dagger_{NN}}{\partial \dot{x}}(\dot{x}_n, \bm{\xi}_n)$
            \State  $\dot{\bm{\xi}}_n = \frac{d D_{NN}^*}{d \dot{\bm{d}}}\left(-\frac{\partial D_{NN}^\dagger}{\partial \bm{\xi}}\left(\dot{x}_n, \bm{\xi}_n\right)\right)$
        \EndFor
        \State Compute Loss $L(w_{D^\dagger}, w_D) = \|f^{RS} - f^{NN}(w_{D^\dagger}, w_D)\|_{{L}^p} / \|f^{RS}\|_{{L}^p}$
        \State Update $w_{D^\dagger}, w_D$ based on the gradient of $L$ w.r.t. $ w_{D^\dagger}, w_D$
    \EndFor
    \State $b \gets b+1$
    \EndWhile
    \State $epoch \gets epoch+1$
\EndWhile
\end{algorithmic}
\end{algorithm}
%
%
%We use Neural Networks to approximate the above potentials within the deeping learning environment PyTorch \cite{paszke2019pytorch}, i.e., 
%\begin{align}
%    W(x) \approx W_{NN}(x; w_W), 
%    D^\dagger(\dot{x}, \bm{\xi}) \approx D^\dagger_{NN}(\dot{x}, \bm{\xi}; w_{D^\dagger}), 
%    D^*\left(\dot{\bm{d}}\right) \approx D^*_{NN}\left(\dot{\bm{d}}, w_{D^*}\right)\label{eq:NNpotentials},  
%\end{align}
%and we call the resulting architecture a Recurrent Neural Operator (RNO) following \cite{BurigedeEric2023, BurigedeMarkovian2023}. 
%
%To find proper $W, D^\dagger$ and $D$ that give $f^{NN}$ similar enough to $f^{RS}$ under a set of rate-and-state parameters, 
%we generate a synthetic dataset of $f^{RS}$'s by prescribing the slip histories $\{V = \dot{x}(t) : t \in [0, T]\}$. 
%We then fit $f^{NN}$'s to $f^{RS}$'s by optimizing over the parameters of $W_{NN}, D^\dagger_{NN}$ and $D_{NN}$, 
%as shown in Figure~\ref{fig:training}. 
%And the loss function we use for training of the potentials is relative $L_p$ error, 
%i.e., 
%\begin{align}
%    w_W^*, w_{D^\dagger}^*, w_D^* = \argmin_{w_W, w_{D^\dagger}, w_D} \frac{1}{N} \sum_{i=1}^N \frac{\|f^{NN}_i(t) - f^{RS}_i(t)\|_{L_p}}{\|f^{RS}_i(t)\|_{L_p}} \label{eq:relativeLpError}. 
%\end{align}

%%%%%%%%%%%%%%%%%%%%%%%%%%%%%%%%%%%%%%%%%%
%\section{Demonstration}
%\label{sec:resultsAndDiscussion}

%%%%%%%%%%%%%%%%%%%%%%%%%%%%%%%%%%%%%%%%%
\subsection{Data and architecture} \label{sec:data}
We consider an example with the following rate-and-state parameters
\begin{align}
    a = 0.011, \
    b = 0.016, \
    f_* = 0.5109, \
    V_* = 1\times 10^{-4}\ \mathrm{m/s}, \
    D_{RS} = 1\times10^{-6}\ \mathrm{m}.
    \label{eq:trainingSeqParams} % f_* &= 0.58, 
\end{align}
Note that $a - b < 0$, and these parameters describe rate-weakening friction that is prone to large accelerations and more challenging to solve numerically.

\begin{figure}[t]
    \centering
    \includegraphics[height=0.3\textheight]{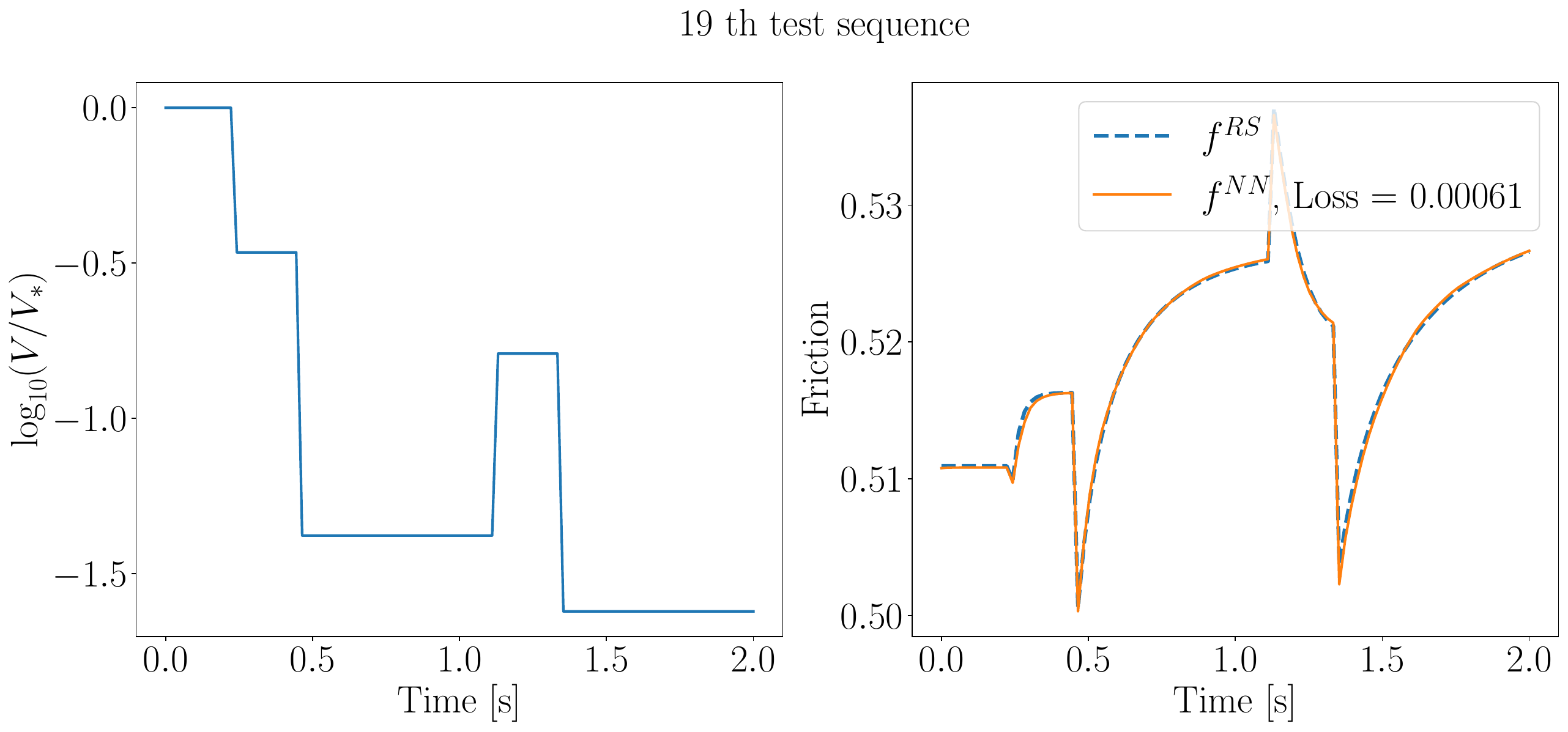}
    \includegraphics[height=0.3\textheight]{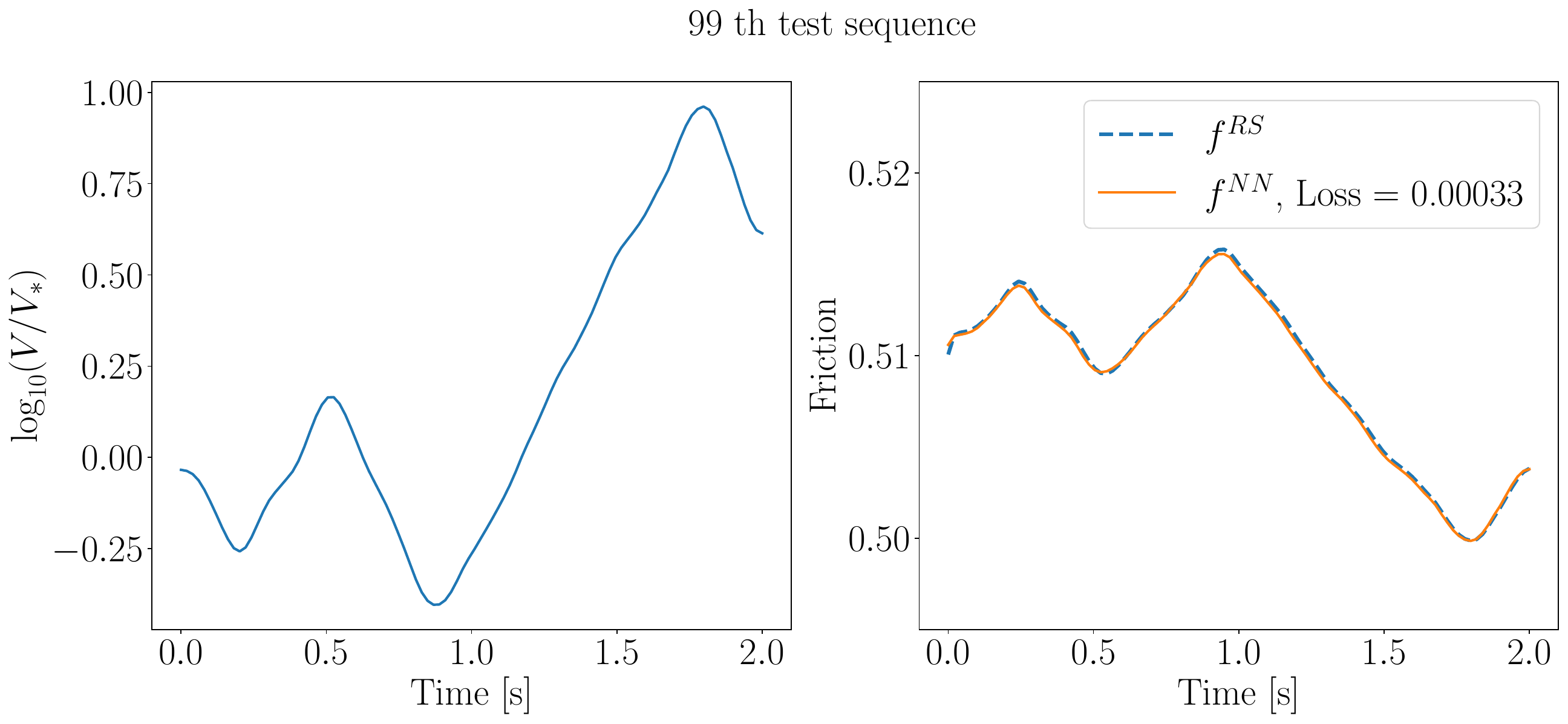}
    \caption{Two typical test examples of data $\{V_i, f_i^{RS}\}$ and results of the trained RNO $\{f_i^{NN}\}$, one from the velocity jump sequence (upper row) and one from a continuous variation sequence (lower row).    Loss is relative $L_2$, (\ref{eq:relativeLpError}), and there is one internal variable in the RNO. }
    \label{fig:19thAnd99thRSNN}
\end{figure}

We take inspiration from both velocity-jump tests typically used to study rate-and-state friction \cite{ruina_slip_1983}, 
as well as continuous variation sequences from the previous studies of RNO \cite{BurigedeEric2023, BurigedeMarkovian2023} to sample velocity trajectories $V_i(t)$ to generate our data.
The velocity-jump sequences are simple functions, i.e., the sum of a finite number of Heaviside functions, 
as shown by the first example in Figure~{\ref{fig:19thAnd99thRSNN}}. 
Note that the prescribed velocity jumps have to be on the logathithmic scale to cause significant changes in $f^{RS}$.  
The continuous-variation sequences
vary continuously with time and change their accelerations at randomly-sampled times, 
as shown by the second example in Figure~\ref{fig:19thAnd99thRSNN}. 
The range of prescribed velocities is set such that $V_{i}(t) / V_* \in [10^{-3}, 10^{1}]$ for both types of sequences. 
% We define 
% $
%     D_* = V_* \cdot 1\ \mathrm{s} = 1\times10^{-6}\ \mathrm{m}. 
% $
%\begin{figure}
%    \centering
%    \includegraphics[height=0.3\textheight]{19thRS.pdf}
%    \includegraphics[height=0.3\textheight]{99thRS.pdf}
%    \caption{Examples of velocity jump $V_i(t)$ (upper, sequence 19), 
%    continuous variation $V_i(t)$ (lower, sequence 99) and their corresponding $f^{RS}$s in the synthetic dataset.}
%    \label{fig:19thAnd99thRS}
%\end{figure}

We generate 480, 160 and 160 sequences for the processes of training, validation and testing respectively, 
using the rate-and-state parameters listed in (\ref{eq:trainingSeqParams}).   For each velocity sequence $V_i(t)$, we solve (\ref{eq:fRS},\ref{eq:AgeingLaw}) to find the corresponding friction force $f_i(t)$ by analytic integration using the package odeint in torchdiffeq \cite{torchdiffeq}.  We then sample the solution at 100 instances with time-step 0.02 secs.  These vectors form our training data.

We train for 100 epoch, with exponential RELU as activation functions and 5 intermediate layers of [1024, 1024, 1024, 1024, 64] neurons at each layer for $D^*_{NN}$, 
as well as 6 intermediate layers of [512, 512, 128, 128, 16, 1024] neurons at each layer for $D^\dagger_{NN}$, 
as chosen by OPTUNA for the hyperparameter-tuning process on training and validation datasets.   The details of the training procedure are given in Algorithm \ref{alg:TrainingOneEpoch}.

%%%%%%%%%%%%%%%%%%%%%%%%%%%%%%%%%%%%%%%%%
\subsection{Comparison of trained potentials with empirical rate-and-state friction}

Recall that we do not specify internal variables a priori, they are learnt as a part of the training process.  However, we have to specify the number (dimension) of internal variables. We train the model with zero, one, and two internal variables, and the resulting errors in comparison with the empirical rate-and-state formulation are shown in Table~\ref{tab:dimXi}.  The results show that it is necessary to have one internal variable to emulate rate-and-state friction, but increasing the number of internal variables beyond one does not lead to better results.  The fact that it suffices to use $\dim(\bm{\xi}) = 1$ makes intuitive sense since the empirical rate-and-state friction that we are trying to match has only one state variable $\theta$.

Figure~\ref{fig:19thAnd99thRSNN} shows the results of two typical examples, one drawn from the velocity-jump sequence and one from the continuous-variation sequence.  Note  that these two sequences are in the test dataset and have not been used for training the potentials.  We observe that the trained RNO with one internal variable is able to reproduce the behavior of rate-and-state friction in these examples.  Table~\ref{tab:dimXi} confirms that the error between $f^{NN}$ and $f^{RS}$ is small across all specimens.

\begin{table}
    \centering
    \begin{tabular}{cccc}
        \hline
        $\dim(\bm{\xi})$ & 0 & 1 & 2 \\
        \hline
        Training error ($L_2$) & 0.18 $\pm$ 0.01 & 0.0004 $\pm$ 0.0004 & 0.0007 $\pm$ 0.0006\\
        Testing error ($L_2$) & 0.18 $\pm$ 0.01 & 0.0005 $\pm$ 0.0004 & 0.0007 $\pm$ 0.0006 \\
        \hline
    \end{tabular}
    \caption{Training and testing relative $L_2$ error for $\dim(\bm{\xi}) = 0, 1, 2$, 
    averaged over 160 test sequences. 
    Error decreases significantly after introducing one hidden variable $\dim(\bm{\xi}) = 1$, 
    while introducing more hidden variables does not further reduce the error. }
    \label{tab:dimXi}
\end{table}

\begin{figure}
    \centering
    \includegraphics[width=0.4\textwidth]{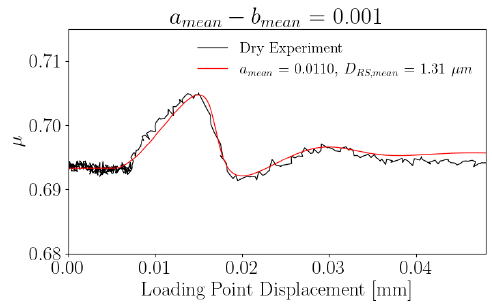}
    \caption{A typical fit of empirical rate-and-state formulation $f^{RS}$ to experimental data $f^{EXP}$; 
    the relative $L_2$ error of $f^{RS}$ against $f^{EXP}$ is 0.0015. 
    Data provided by Taeho Kim \cite{Kim2025Modeling}.}
    \label{fig:RSVsExp}
\end{figure}

We now seek to provide some meaning to the magnitude of errors in Table~\ref{tab:dimXi}. We note that the error of fitting the RNO to the empirical rate-and-state friction (Table~\ref{tab:dimXi}) is much smaller than the typical error of fitting the empirical rate-and-state friction model to actual experimental data.  Figure~\ref{fig:RSVsExp} (Kim, 2025 \cite{Kim2025Modeling}) shows an experimental sequence $f^{EXP}$ and the best fit with the empirical rate-and-state formulation $f^{RS}$: the relative $L_2$ error is $0.0015$.  This is an order of magnitude higher than the error of the RNO relative to the empirical rate-and-state formulation, and hence the errors in Table~\ref{tab:dimXi} with one internal variable are meaningfully small.  Further, it follows that the RNO architecture proposed here would fit the experimental data with an accuracy similar to that of the empirical rate-and-state formulation.  We are unable to train the RNO directly from experiments because training requires significant amounts of data, and experimental data is sparse. 

%
%
%
%Due to limited access to experimental data with the same rate-and-state friction properties, 
%we cannot compare the error of $f^{RS}$ and $f^{NN}$ both fitted to 
%experimental sequences $f^{EXP}$. 
%We here include a typical rate-and-state fitted experimental sequence from (Kim et al., in preparation, 2024), 
%which is shown by Figure~\ref{fig:RSVsExp}. 
%The best fit rate-and-state sequence achieves an relative $L_2$ error of $0.0268$, 
%which is two orders of magnitude higher than the average relative $L_2$ of fitting $f^{NN}$ to $f^{RS}$. 
%This implies that the fitting error between $f^{NN}$ and $f^{RS}$ is negligible compared with fitting $f^{RS}$ to noisier $f^{EXP}$, 
%and thus $f^{NN}$ has comparable ability to explain the history dependencies in the empirical observations. 
%
%

%%%%%%%%%%%%%%%%%%%%%%%%%%%%%%%%%%%%%%%%%
\subsection{Properties of the trained potential formulation}

\begin{figure}
    \centering
 \begin{subfigure}{2.5in}
    \includegraphics[width=2.5in]{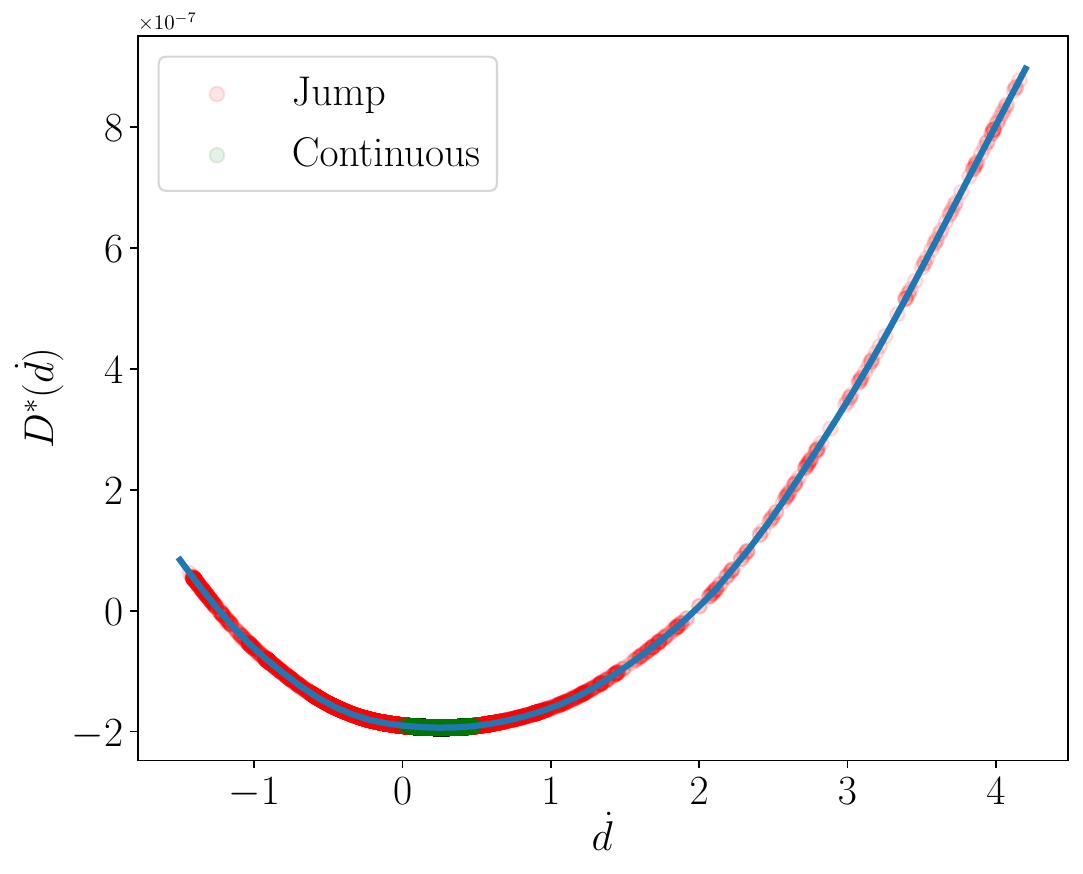} 
    \caption{}
 \end{subfigure}
 \hfill
 \begin{subfigure}{3.5in}
    \includegraphics[width=3.5in]{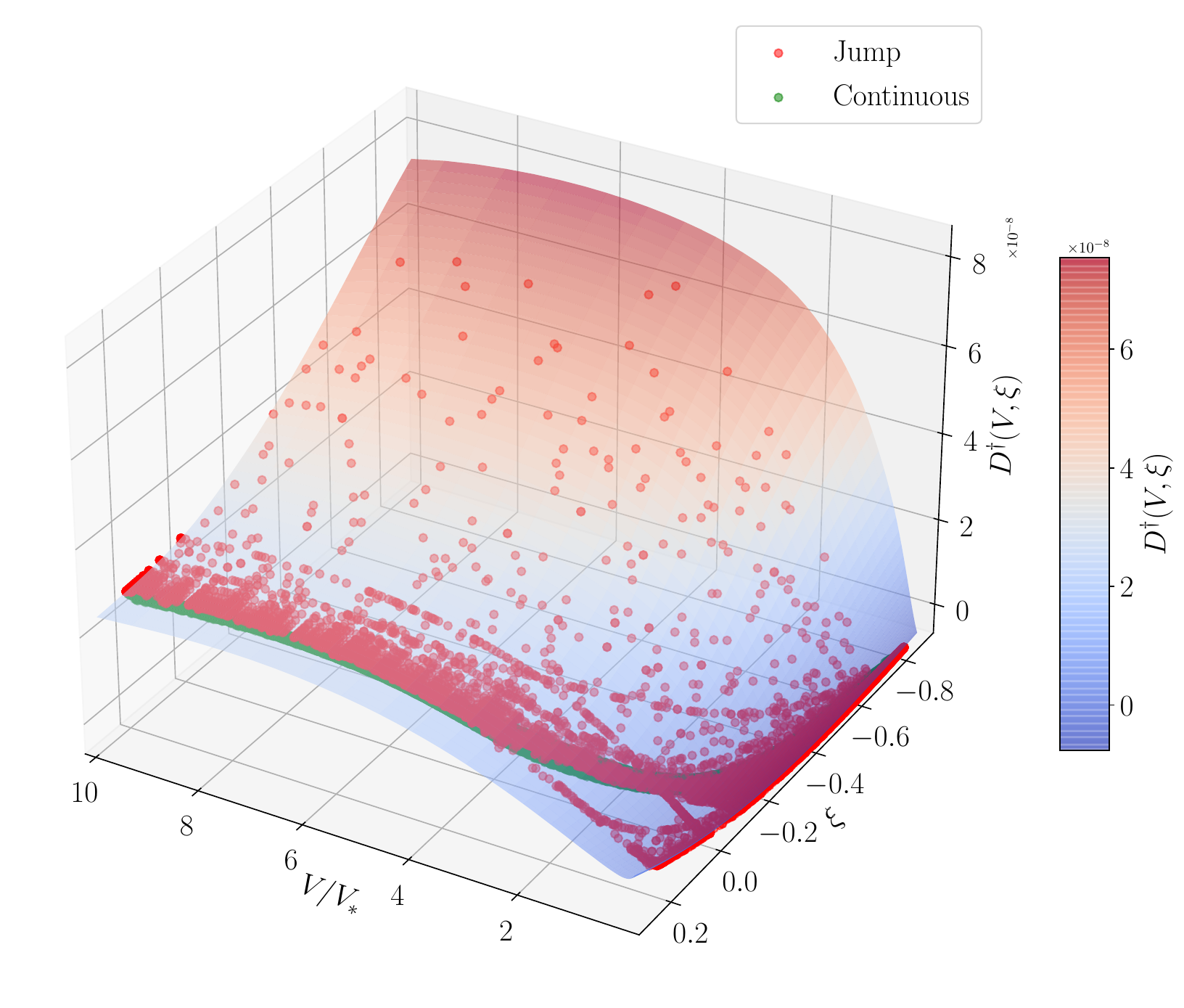}
    \caption{}
\end{subfigure}
    \caption{Learnt potentials (a) $D^*(\dot{d})$ and (b) $D^\dagger$. 
    $D^*$ is convex, but $D^\dagger$ is not convex in $(\dot{x}, \xi)$.
    The red dots show the trajectories of velocity-jump dataset, 
    while the green dots show the trajectories of continuous variation dataset.}
    \label{fig:WAndD}
\end{figure}

\begin{figure}[t]
    \centering
    \includegraphics[width=0.9\textwidth]{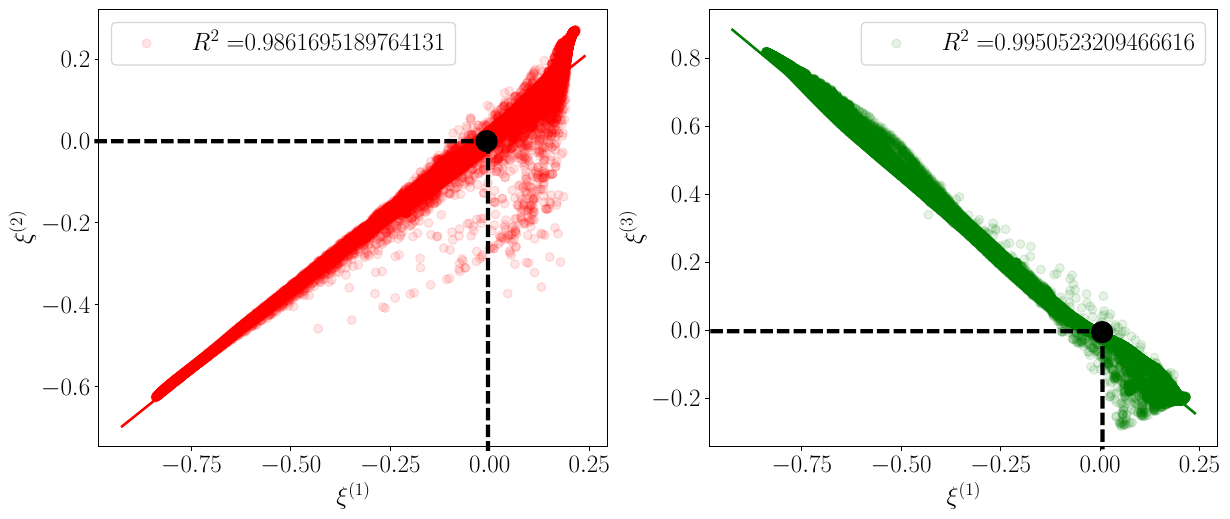}
    \caption{Comparison of the internal variables obtained from three models trained using different datasets all corresponding to the same underlying empirical rate-and-state formulation with the same parameters.}
    \label{fig:nonuniqueXis}
\end{figure}

Figure~\ref{fig:WAndD} plots the learnt potentials for the case with a single internal variable.  $D^*$ is convex in its argument $\dot{d}$.    However, $D^\dagger(\dot{x}, \xi)$ is not convex because $\partial^2 D^\dagger / \partial \dot{x}^2 < 0$.   This corresponds to the rate-weakening nature of the underlying data.

%\begin{figure}[htb!]
%    \centering
%    \includegraphics[height=0.5\textheight]{Trial0216_combined_800_D_dagger_normal.pdf}
%    \caption{Learnt $D^\dagger(\dot{x}, \bm{\xi})$, 
%    $D^\dagger$ is not convex in $(\dot{x}, \bm{\xi})$.
%    The red dots show the trajectories of velocity-jump dataset, 
%    while the green dots show the trajectories of continuous variation dataset.}
%    \label{fig:Ddagger}
%\end{figure}

%In summary, 
%we find that of the three learnt potentials, 
%$W$ is linear in $x$, 
%consistent with material-frame indifference; 
%$D^*$ is convex in $\dot{d}$, 
%consistent with its definition by Legendre transform; 
%while $D^\dagger$ is not convex in $(\dot{x}, \bm{\xi})$. 
%However, 
%one can still achieve convexity of $J(x, \bm{\xi})$ in (\ref{eq:Jfunctional}) with an upper bound constraint on $\Delta t$, 
%and thus it is legitimate to write (\ref{eq:JfunctionalMin}) as a convex minimization problem. 

Let us consider the uniqueness of the internal variable $\bm{\xi}$.  Since it is learnt from the data, and since neural network training often leads to local minima, it is possible that two separate instances of training from the same data can lead to two different models and internal variables.   However, an underlying assumption in our potential formulation is that $D$ is only a function of $\dot{\bm{\xi}}$.  In such a setting, under some hypothesis, it is possible to show (see \cite{Raj} for the rigorous argument) that the internal variable is unique up to an affine transformation.  Briefly, if the behavior is described by two separate internal variables $\bm{\xi}$ and $\eta$ (and two separate sets of potentials $\hat{D}^\dagger, \hat{D}$), there must be a function $g$, with $g$ and $g^{-1}$  smooth, such that $\eta = g(\bm{\xi})$.  We can show that 
\begin{align}
    \hat D_\eta\left(g'(\bm{\xi})\dot{\bm{\xi}}\right) = D_\xi \left(\dot{\bm{\xi}}\right) \label{eq:uniqueXi}, 
\end{align}
which implies that $g'(\bm{\xi})$ is a constant under suitable hypothesis.  Thus, the internal variable $\bm{\xi}$ is unique up to affine transformations. 

We demonstrate this uniqueness of the internal variable by training the model on three datasets generated using the same rate-and-state friction model. We then apply these three models on the same test dataset and obtain the trajectories of the internal variables.   Figure~\ref{fig:nonuniqueXis} compares the internal variables for the three distinctly trained models.  The linear regression coefficient is $>0.98$ between the internal variables of the three models, and thus the different internal variables are indeed unique up to linear transformations.

%%%%%%%%%%%%%%%%%%%%%%%%%%%%%%%%%%%%%%%%%
\section{Application: Spring-slider system}
\label{sec:sss}

In this section, we apply the trained RNO with potential formulation to an application.  We recall the spring-slider model in Section \ref{sec:ss}.  We then show in Section \ref{sec:var} that the potential formulation with implicit forward Euler time discretization leads to a variational problem that enables stable updates.  We then assess the RNO on two complementary questions.  The first question, addressed in Section \ref{sec:err}, concerns the accuracy of the RNO trained using velocity histories in Section \ref{sec:data} in a fully resolved variable time-step calculation.  The second question addressed in Section \ref{sec:stab} assesses the stability of the RNO, which is based on a potential formulation, in a fixed time-step calculation.

%We now apply the potential formulation to simulate the motion of the spring-slider system shown in Figure~\ref{fig:springslider} and assess two important issues.  First, the velocity profiles that we use to train the neural network potentials are not chosen from an application, and therefore it is important to assess the ability of the trained model to replicate the behavior of the empirical rate-and-state friction in an application.  Second, we examine the numerical performance of the potential formulation with the trained potentials.

%%%%%%%%%%%%%%%%%%%%%%%%%%%%%%%%%%%%%%%%%
\subsection{Spring-slider system} \label{sec:ss}

\begin{figure}
    \centering
    \includegraphics[width=0.4\textwidth]{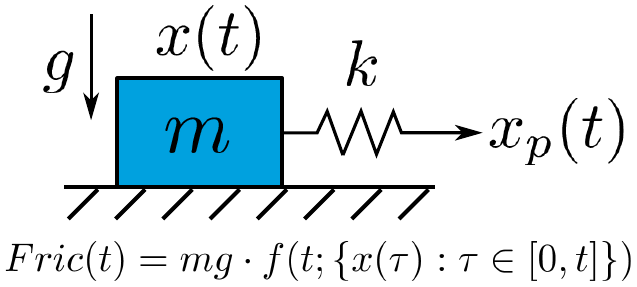}
    \caption{Example: spring slider under displacement-control driving force.}
    \label{fig:springslider}
\end{figure}

We consider the spring-slider system shown in Figure \ref{fig:springslider} in which a slider of mass $m$ is connected through a linear spring of stiffness $k$ to a loading point moving with displacement $x_p(t)$.   The motion of the loading point applies a force to the block through the spring, and the motion of the block is resisted by the friction force $mgf$ below the block, where $g$ is the acceleration due to gravity.  The balance of forces on the block involves the inertial, spring, and friction forces, resulting in the following equation of motion:
\begin{align}
    m\Ddot{x} - k(x_p(t) - x(t)) + mgf  &= 0. \label{eq:springslidergeneral}
\end{align}
For cases with the empirical rate-and-state formulation, the friction coefficient $f$ is given by equations (2-3) \cite{Dieterich1978, ruina_slip_1983}. For cases with the potential formulation developed in this study, $f$ is given by equations (\ref{eq:fpot}, \ref{eq:ComputeXiDot}).

This simple system is known to have extremely complex solutions.  For example, when a steady loading velocity $\dot{x}_p$ is imposed on the spring slider system, there is a critical spring constant $k_{crit}$ below which the steady sliding is stable to perturbations and above which the sliding can develop stick-slip behavior \cite{rice_stability_1983, Gu_Rice_1984}.

%%%%%%%%%%%%%%%%%%%%%%%%%%%%%%%%%%%%%%%%%
\subsection{Variational update  of the potential formulation} \label{sec:var}

In the potential formulation developed in this study, the equations of motion of the spring-slider system are:
\begin{align}
    m\Ddot{x} - k(x_p(t) - x(t)) + mg  \frac{\partial \widetilde D^\dagger}{\partial \dot{x}}&= 0 \label{eq:springsliderEOM1} \\
    \frac{d D}{d \dot{\boldsymbol{\xi}}}(\dot{\bm{\xi}}) + \frac{\partial \widetilde D^\dagger}{\partial \boldsymbol{\xi}}(\dot{x}, \bm{\xi}) &= 0 \notag
\end{align}
  
In the implicit time (forward Euler) discretization of this system, given $\{x_{n-1},x_n,\bm{\xi}_n\}$, we aim to find $\{ x_{n+1}, \bm{\xi}_{n+1}\}$ such that:
\begin{align}
    m \left(\frac{x_{n+1} - 2 x_n + x_{n-1}}{\Delta t ^2} \right)- k(x_p(t_{n+1}) - x_{n+1} ) 
     +  mg  \frac{\partial \widetilde D^\dagger}{\partial \dot{x}}\left(\frac{x_{n+1} - x_n}{\Delta t}, \bm{\xi}_{n+1}\right) = 0, \label{eq:springsliderEOM2} \\
    \frac{d D}{d \dot{\boldsymbol{\xi}}}\left(\frac{\bm{\xi}_{n+1} - \bm{\xi}_n}{\Delta t}\right)  + \frac{\partial \widetilde D^\dagger}{\partial \boldsymbol{\xi}}\left(\frac{x_{n+1} - x_n}{\Delta t}, \bm{\xi}_{n+1}\right) = 0. \notag
\end{align}
This can be written as a variational problem:
\begin{align}
    x_{n+1}, \bm{\xi}_{n+1} &= \argmin_{x, \bm{\xi}} J(x, \bm{\xi}; \Delta t, x_n, \bm{\xi}_{n}), \label{eq:JfunctionalMin} 
\end{align}
where the system potential is:
\begin{align}
    J(x, \bm{\xi}; \Delta t, x_n, \bm{\xi}_n) &= E^{sp}(x) + E^{in}(x) + W(x) + \Delta t \widetilde D^\dagger\left(\frac{x - x_n}{\Delta t}, \bm{\xi}\right) + \Delta t^2 D\left(\frac{\bm{\xi} - \bm{\xi}_n}{\Delta t}\right), \label{eq:Jfunctional}
\end{align}
with spring and the inertia potentials given by:
\begin{align}
    E^{sp}(x) = \frac{1}{2}\left(x(t) - x_p(t)\right)^2, \quad \quad 
    E^{in}(x) = \frac{1}{2}\left(\frac{x - 2 x_n + x_{n-1}}{\Delta t}\right)^2 \label{eq:Ein}, 
\end{align}
respectively.  This problem is well-posed if $J$ is convex in $(x, \bm{\xi})$.  Convexity of $J$ is trivial if $D^\dagger, D$ are both convex in $(x, \dot{x}, \bm{\xi})$.  However,  $D^{\dagger}$ may not be convex if the friction coefficient decreases with slip rate,
\begin{align}
    \frac{\partial f}{\partial \dot{x}} = \frac{\partial^2 \widetilde D^\dagger}{\partial \dot{x}^2} < 0 \label{eq:nonConvexDdagger}. 
\end{align}
Indeed, we see this in our trained potential in Figure \ref{fig:WAndD}.

In such cases, $J$ can still be convex in $(x, \bm{\xi})$ if time step $\Delta t$ is small enough, and this limits the time-step in a numerical implementation.  Studying the Hessian of $J$, we conclude that the time step $\Delta t$ should satisfy the following conditions:
\begin{align}
    1 \cdot m + \Delta t \left(\frac{\partial^2 D^\dagger}{\partial \dot{x}^2}\right) + O(\Delta t^2)  &\ge 0, \label{eq:PSD1}\\
     1 \cdot m \frac{d^2D}{d\dot{\xi}^2} 
    + \Delta t \left(\frac{\partial^2 D^\dagger}{\partial \dot{x}^2} \frac{d^2D}{d\dot{\bm{\xi}}^2} + m \frac{\partial^2 D^\dagger}{\partial \bm{\xi}^2}\right) 
    \notag \\
     + \Delta t^2 \left[\left(k + \frac{d^2W}{dx^2}\right)\frac{d^2D}{d\dot{\bm{\xi}}^2} + \frac{\partial^2 D^\dagger}{\partial \dot{x}^2}\frac{\partial^2 D^\dagger}{\partial \bm{\xi}^2}-\frac{\partial^2D^\dagger}{\partial \dot{x} \partial \bm{\xi}}\right]  
     + \Delta t^3 \left[\left(k + \frac{d^2W}{dx^2}\right)\frac{\partial^2D^\dagger}{\partial \bm{\xi}^2}\right] & \ge 0. \label{eq:PSD2}
\end{align}
We can select $\Delta t$ to satisfy these equations since $m>0, k>0, d^2 D / d \dot{\bm{\xi}}^2 > 0, d^2 D^\dagger / d \bm{\xi}^2 > 0$, and $\partial^2 D^\dagger / \partial \dot{x}^2$ is bounded from below.  While this establishes the conditions for stability, it is difficult to asses the limits of stability explicitly in the RNO formulation since we do not know the potentials explicitly.

%%%%%%%%%%%%%%%%%%%%%%%%%%%%%%%%%%%%%%%%%
\subsection{Error of pre-trained RNO and transfer learning} \label{sec:err}

We now assess the accuracy of the RNO that was trained using the velocity histories in Section \ref{sec:data}.  Recall that these velocity histories are not sampled from an actual application.  So we assess the accuracy of the RNO against the RS friction model in the spring-slider system.  

%it is a natural question to compare the results of the trained RNO and RS friction models in an actual application.  We do so in Section \ref{sec:err} by solving a spring slider system using a variable time integration scheme for both the RNO and RS friction model.  We find that the accuracy in the application is smaller than that with velocity histories in Section \ref{sec:data}.  This indicates that either the formulation or the sampled velocities are deficient.  To remedy this, we use transfer learning (initialize with the parameters trained on velocity histories in Section \ref{sec:data} and then 

We select 200 imposed trajectories $x_p(t)$ that result in both steady and unsteady behavior. They involve loading at a constant velocity $\dot{x}_p$ for some interval of time and then rapidly changing this velocity to another constant value. The chosen loading trajectories result in rapid decaying oscillation in sliding velocity, the behavior significantly different from the training set, but not in a fully developed stick-slip in which the sliding velocity changes over several orders of magnitude, an instability possible in the system \cite{rice_stability_1983, Gu_Rice_1984}. We solve both the RNO and empirical RS models using a variable time-step method (implicit for the RNO and explicit for the RS) to control numerical errors at each time-step.

\begin{figure}
    \centering
    \includegraphics[width=0.9\textwidth]{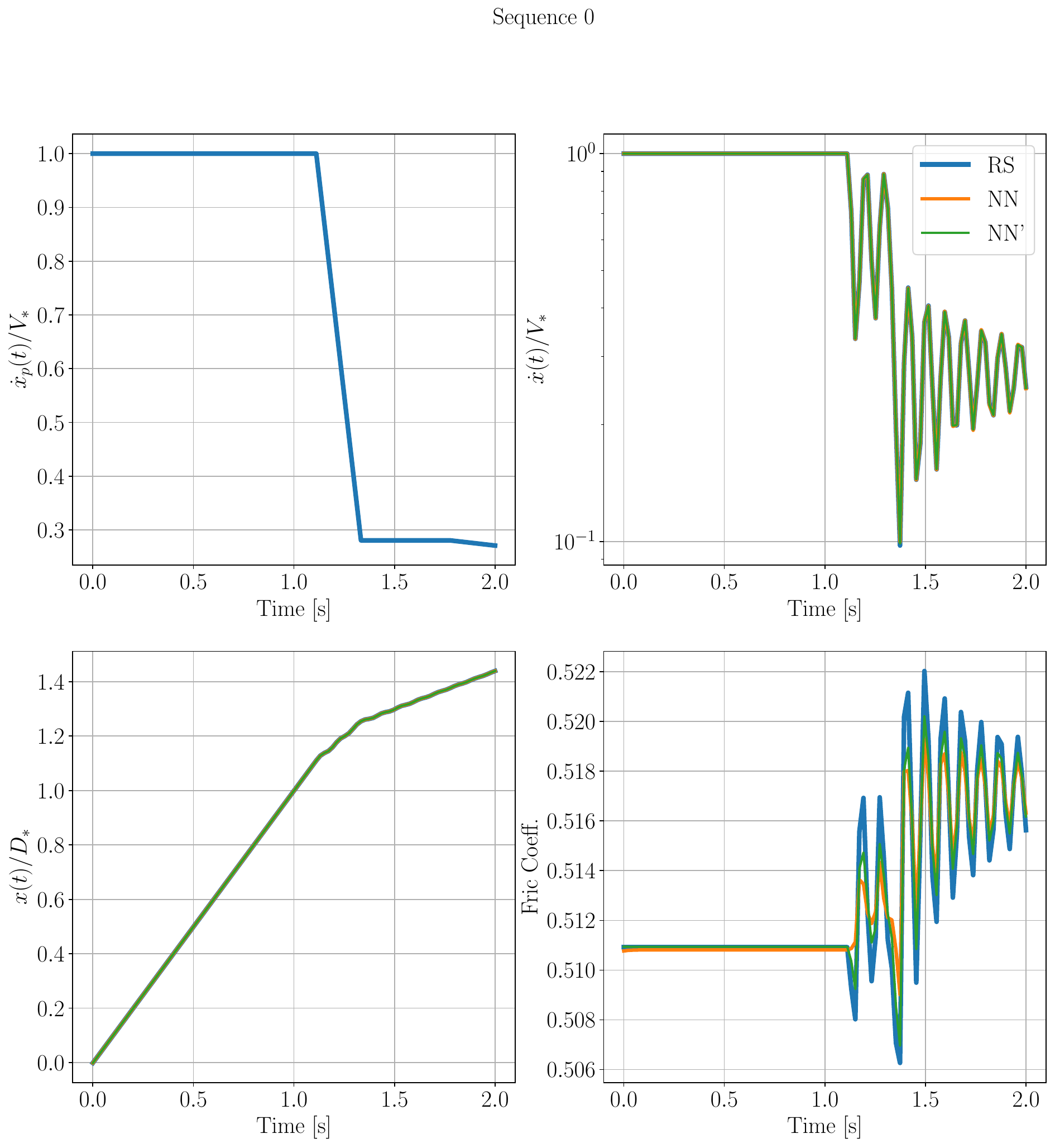}
    \caption{An example of spring-slider simulations with the empirical rate-and-state friction (blue lines), the trained NN potentials (orange lines), and the NN potentials further trained on spring-slider sequences (green lines).  The velocity of the loading point is given in the top left panel, while the displacement, velocity, and friction coefficient of the block is given in the other three panels.}
    \label{fig:SSseq8}
\end{figure}

The typical results for an unsteady sequence involving oscillating slip-velocity behavior are shown in Figure \ref{fig:SSseq8}.  We observe that the solution with the trained potential (labeled NN) reproduces the solution with the empirical rate-and-state formulation (labeled RS) well overall, with some differences in the fluctuations of the friction coefficient.   The average relative $L_2$ errors in the position (slip), velocity, and friction coefficient are given in Table \ref{tab:NNvsNNPrime} (column labeled NN).  We see that the error is small, but larger than those in Table \ref{tab:dimXi} that uses the same class of velocity profiles as the training.  Thus, while training on independently generated velocity histories is still effective on out-of-distribution tests, it has larger errors.

To confirm that the larger errors are due to the out-of-distribution tests, and not due to deficiencies of the formulation, we create a new data set using 200 velocity profiles generated by the solution of the spring-slider system using the RS model.  We use transfer learning following \cite{zhang_2024}: we initialize the neural networks using the parameters obtained in Section \ref{sec:data}, and then further optimize it over this new data set. We find that this leads to significantly lower error (Figure \ref{fig:SSseq8}, Table \ref{tab:NNvsNNPrime}), confirming that the larger errors were due to the lack of adequate data and not a deficiency of the formulation.

We use these enhanced potentials in the following section.

%We notice that since the training dataset of the potentials does not include these spring-slider sequences, the error between the potential friction and the original rate-and-state friction is large.  To resolve this,  we generate $200$ sequences from solving the spring-slider prowblem with rate-and-state friction and different loadings, and further train our potentials on these $200$ sequences for $400$ epochs.   Indeed the further-trained NN potentials (denoted as NN') reduce the error between $f^{NN'}$ and $f^{RS}$ when solving spring-slider sequences. 
%
%Table~\ref{tab:NNvsNNPrime} shows that after further training the potentials on spring-slider solutions by rate-and-state friction, 
%the relative $L_2$ error on $x(t), \dot{x}(t), f(t)$ decreases by more than $50\%$. 
%Figure~\ref{fig:SSseq8} plots an example spring-slider sequence. 
%It is clear that further trained NN' agrees better with the solution obtained by the original rate-and-state friction. 
%Another example sequence is shown by Figure~\ref{fig:SSseq9}.
%The results and discussions next will all be based on the solution of NN'. 
%

 \begin{table}
    \centering
    \begin{tabular}{ccc}
        \hline
        Solution term & NN & NN' \\
        \hline
        $x(t)$ & $(1.9 \pm 1.8)\times 10^{-5}$ & $(5.4 \pm 4.4 )\times 10^{-6}$\\
        $\dot{x}(t)$ & $(2.7 \pm 3.6)\times 10^{-4}$ & $(1.6 \pm 2.2)\times 10^{-4}$\\
        $f(t)$ & $0.030 \pm 0.024$ & $0.016 \pm 0.016$\\
        \hline
    \end{tabular}
    \caption{Testing relative $L_2$ error for the original potentials only trained on velocity-jump and continuous variation dataset (NN), 
    updated potentials further trained on 200 spring-slider like dataset (NN') 
    averaged over 10 test spring-slider sequences. }
    \label{tab:NNvsNNPrime}
\end{table}

%%%%%%%%%%%%%%%%%%%%%%%%%%%%%%%%%%%%%%%%%
\subsection{Comparison of numerical stability} \label{sec:stab}

We now address the complementary issue of numerical performance of the RNO with the potential formulation.  Recall from Section \ref{sec:var} that the potential formulation of the RNO enables a variational update of the implicit-time discretization.  We expect that this should lead to better stability and accuracy at a fixed time step.

We choose 77 sets of applied displacements $x_p(t)$ at constant velocity, with the loading velocity and spring constant selected randomly from a range that encompasses both the steady and unsteady sliding regimes.  For each set, we solve both the RNO and RS models using both explicit (4th order Runge Kutta) and implicit (Adams) solvers (cf.\ ODE solver package torchdiffeq \cite{torchdiffeq}) with fixed time steps ranging from $2^{-13.5} (8.63 \times 10^{-5})$ to $2^{-11} (4.88 \times 10^{-4}) $ seconds.

The results on stability are shown in Table \ref{tab:NaNRatioSpringSliderRsVsNNRespective}.  We find that the RNO provides a stable solution for each of the 77 cases with both implicit and explicit time discretization at all choices of time-step.  This is not true for the RS friction formulation, for which most simulations become unstable and diverges (returning `nan' or `not a number') for the implicit time discretization and almost half do not converge for the explicit approach, even for the finest time step (Table~\ref{tab:NaNRatioSpringSliderRsVsNNRespective}). Indeed, decreasing the time step to $2^{-19} \approx 10^{-6}$ seconds still does not change the non-convergent fraction of implicit solves with the empirical rate-and-state friction.  This confirms the anticipated superior stability of the RNO with the potential formulation.

%The poor stability of the implicit solver with the RS friction is not surprising, since it does not have an associated variational formulation.   The fact that the potential friction formulation works well with the implicit solver is consistent with (\ref{eq:JfunctionalMin}) being a convex minimization problem. 

%\hl{Do you have comparison of computation wall-clock times?}

\begin{table}
    \centering
    \begin{tabular}{cccccccc}
        \hline
        $\Delta t$ [s] & $2^{-13.5}$ & $2^{-13.0}$ & $2^{-12.5}$ & $2^{-12.0}$ & $2^{-11.5}$ & $2^{-11.0}$ \\
        \hline
        NN, implicit & 0.000 & 0.000 & 0.000 & 0.000 & 0.000 & 0.000 \\
        NN, explicit & 0.000 & 0.000 & 0.000 & 0.000 & 0.000 & 0.000 \\
        RS, implicit & 0.506 & 0.571 & 0.623 & 0.623 & 0.675 & 0.727 \\
        RS, explicit & 0.455 & 0.455 & 0.455 & 0.455 & 0.455 & 0.455 \\
        \hline
    \end{tabular}
    \caption{Ratio of simulations that become unstable and do not converge with the potential friction formulation (NN), empirical rate-and-state friction (RS) and implicit, explicit solvers.}
    \label{tab:NaNRatioSpringSliderRsVsNNRespective}
\end{table}

\begin{figure}
    \centering
    \includegraphics[width=0.7\textwidth]{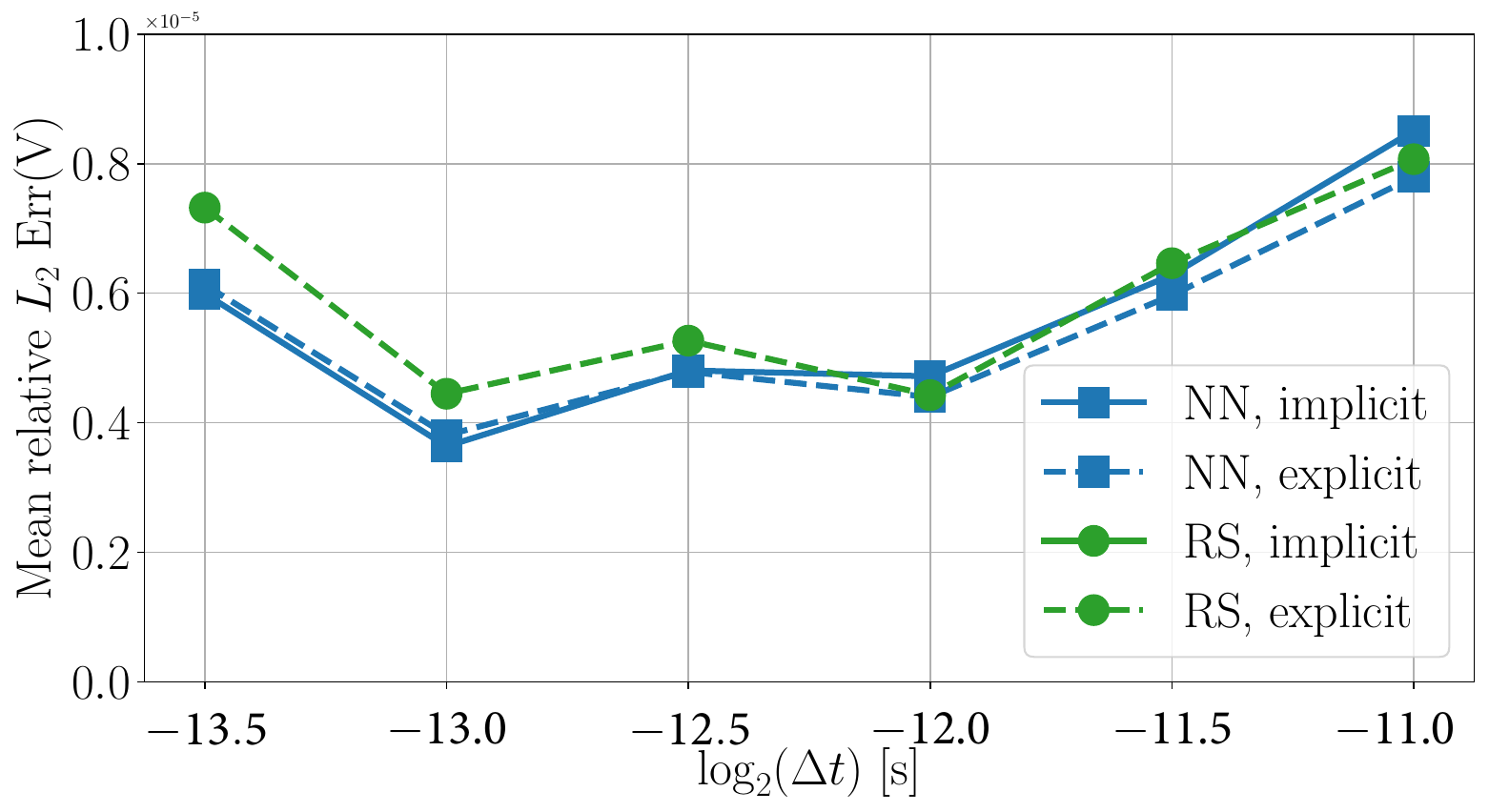}
    \caption{Growth of relative $L_2$ error in $\dot{x}(t)$ as $\Delta t$ increases.}
    \label{fig:ErrGrowthDt}
\end{figure}

The stability of solutions is often related to the error and its growth with time-step.  To probe this, we turn to the growth of error as the fixed time step $\Delta t$ increases.   Since we do not have analytical solutions for the sequences, the $L_2$ error is computed against the solution with  $\Delta t = 2^{-14}\ \mathrm{s}$.   In the case of the RS formulation with explicit time discretization we only consider those cases where are able to solve the equations for all time steps.  We do not consider simulations with the RS formulation and implicit time discretization as so few of those simulations converge.  The $L_2$ error in the velocity $\dot{x}$ averaged over 77 sequences is shown in Figure \ref{fig:ErrGrowthDt} with details in Table~\ref{tab:L2ErrorSpringSliderRsVsNNRespective}.   We observe that the error grows mildly with increasing time step, and the error is on the order of $10^{-5}$. In other words, when the solution remains stable, the error remains relatively controlled even with the increase of time-step in the range studied.

In summary, we conclude that the potential formulation provides superior numerical stability.

%
%Next, 
%we check the growth of error as $\Delta t$ increases for those sequences that (NN, implicit), (NN, explicit) and (RS, explicit) can all solve. 
%Since we do not have analytical solutions for the sequences, 
%Error is computed against the solve with the finest $\delta t = 2^{-14}\ \mathrm{s}$, for each (model, ex/implicit) pair. 
%Figure~\ref{fig:ErrGrowthDt} shows that the error in $\dot{x}(t)$ is on the order of $10^{-5}$, 
%while further increasing $\Delta t$ would result in (RS, explicit) not solving some of the sequences listed here. 
%We conclude that within this range of $\Delta t$ such that all the three pairs can solve these sequences, 
%their error growth is comparable and small, 
%since the fitting error between potential formulated friction and rate-and-state friction is already on the order of $10^{-4}$. 
%For the sequences that (RS, explicit) cannot solve, 
%further decreasing the time step to $\Delta t = 2^{-19}\approx 10^{-6}\ \mathrm{s}$ still will not solve them, 
%while further decreasing $\Delta t$ is of little practical value since that is close to the precision of float tensors on GPUs. 

\begin{table}
    \centering
    \begin{tabular}{ccccccc}
        \hline
        $\Delta t$ [s] & $2^{-13.5}$ & $2^{-13.0}$ & $2^{-12.5}$ & $2^{-12.0}$ & $2^{-11.5}$ & $2^{-11.0}$ \\
        \hline
        \multicolumn{7}{c}{Mean Relative L$_2$ error}\\
        \hline
        NN, implicit & 5.993e-06 & 3.636e-06 & 4.807e-06 & 4.716e-06 & 6.282e-06 & 8.508e-06 \\
        NN, explicit & 6.130e-06 & 3.808e-06 & 4.786e-06 & 4.397e-06 & 5.968e-06 & 7.795e-06 \\
 %       RS, implicit & nan & nan & nan & nan & nan & nan \\
        RS, explicit & 7.321e-06 & 4.447e-06 & 5.267e-06 & 4.426e-06 & 6.464e-06 & 8.069e-06 \\
        \hline
         \multicolumn{7}{c}{Standard Deviation in Relative L$_2$ error}\\
        \hline
        NN, implicit & 5.799e-06 & 5.390e-06 & 5.575e-06 & 7.069e-06 & 6.384e-06 & 1.033e-05 \\
        NN, explicit & 6.241e-06 & 5.766e-06 & 5.844e-06 & 6.887e-06 & 6.572e-06 & 9.639e-06 \\
 %    RS, implicit & nan & nan & nan & nan & nan & nan \\
        RS, explicit & 9.601e-06 & 6.886e-06 & 5.541e-06 & 5.597e-06 & 6.845e-06 & 1.017e-05 \\
        \hline
   \end{tabular}
    \caption{Mean and standard deviation of relative $L_2$ error in $\dot{x}(t)$ averaged over 77 sequences for the potential friction formulation (NN), empirical rate-and-state friction (RS), and implicit, explicit solvers.}
    \label{tab:L2ErrorSpringSliderRsVsNNRespective}
\end{table}

\section{Conclusions}
\label{sec:conclusions}

%\hl{KB: Not touched this section}
We have developed a potential friction formulation that mimics widely-used empirical rate-and-state friction laws by constructing the potentials using recurrent neural operators (RNOs) and training them on data generated using the empirical rate-and-state formulations.  Testing the potential formulation on time histories of slip rates similar to the training set shows that the RNO architecture with one internal variable fits the empirical rate-and-state friction quite well.  Specifically, for a prescribed slip-rate evolution, the average error in the resulting friction coefficient between the newly developed potential formulation and the empirical rate-and-state one is much smaller than the error of fitting the experimental results with the empirical formulation. We find that increasing the number of internal variables beyond one does not lead to better results, consistent with having one state variable in the empirical formulation.  We demonstrate that the internal variable of the potential formulation is unique up to affine transformations. 

We have applied the developed potential formulation to simulate the motion of the spring-slider system and showed that the trained model can replicate the behavior of the empirical rate-and-state friction in an application that results in slip-rate histories quite different from the training set. At the same time, the error is large in this case; training the potentials further on the slip-rate histories from the spring-slider simulations reduces the error.  

Furthermore, we find that the potential friction formulation indeed facilitates both implicit and explicit solutions of initial value problems with rate-and-state frictional interfaces. Although one of the trained potentials is not convex, the implicit discretization of the equations can still be written as a convex minimization problem, provided the time steps are small enough. With the potential friction formulation, the solution to the spring-slider problem converges for both implicit and explicit schemes and suitably selected time steps, for all loading scenarios considered.  This is a clear improvement over the empirical rate-and-state friction, the solution for which cannot be numerically determined in more than half of the simulated cases with the implicit solver, and nearly half with the explicit solver, even when the time step is reduced 256 times. Within the sequences that both explicit rate-and-state and implicit potential friction can solve, they achieve similar accuracy for the same time steps. 

With all the advantages of the potential formulation for rate-and-state friction, one drawback is the large amount of data required to train the RNO potentials. Compared to the empirical rate-and-state friction law, the potentials have many more parameters from their Neural Network structure, and hence hundreds of sequences are required to avoid over-fitting and achieve good performance.  Future directions include finding closed-form approximations to the learnt potentials so that taking their gradients can be done more efficiently, considering larger range of variations in slip rates, and fitting the potential formulations directly to experimental measurements and/or to the combination of experimental measurements and data from empirical friction laws. 

%\newpage
%\appendix
%\section{Supplementary figures and tables}
%\begin{figure}[htb!]
%    \centering
%    \includegraphics[width=0.9\textwidth]{SS_seq9_0216_0521SS_combined_800.pdf}
%    \caption{An example sequence of spring-slider solution with original rate-and-state friction, 
%    NN potentials, 
%    and NN potentials further trained on spring-slider sequences}
%    \label{fig:SSseq9}
%\end{figure}
%

\section*{Acknowledgement}
We gratefully acknowledge the financial support of the National Science Foundation through grant NSFGEO-NERC 2139331 to NL and the Office of Naval Research through MURI grant N00014-23-1-2654 to KB.

%Bibliography
%\newpage

\end{document}